%% file: ppfupeah.tex
\documentclass[aps,prd,twocolumn,groupedaddress,nofootinbib]{revtex4}
\pdfoutput=1

\usepackage{tikz}
\usetikzlibrary{positioning,shapes,shadows,arrows}

\usepackage {natbib}
\usepackage {graphicx}
\usepackage {color}

\usepackage{amsmath, amsfonts, amssymb}
\usepackage{hyperref}

\usepackage[caption=false]{subfig}

\input{tag.tex}
\newcommand{\dcc}{LIGO-P1400057-v2}
\newcommand{\AUTSTRING}{M.\ Shaltev, P.\ Leaci$^{*}$, M.\ A.\ Papa$^{*}$,\ R.\ Prix}

\newcommand{\ltitle}{
  Fully coherent follow-up of continuous gravitational-wave candidates:
  an application to Einstein@Home results
}
\newcommand{\stitle}{Coherent follow-up of Einstein@Home candidates}

\hypersetup{
  pdftitle={\ltitle},
  pdfauthor={\AUTSTRING},
  pdfkeywords={\$\commitID-\commitSTATUS{} \$}
}

\input{ppfupeahdefs.tex}

\begin{document}

\title[\stitle]{\ltitle}
\author{\AUTSTRING}
\affiliation{Albert-Einstein-Institut, Callinstr.\ 38, 30167 Hannover, Germany\\$^{*}$Albert-Einstein-Institut, Am M\"uhlenberg\ 1, 14476 Potsdam-Golm, Germany}
\date{\commitDATE\\\mbox{\dcc}\\\commitIDshort-\commitSTATUS}


\begin{abstract}
We characterize and present the details of the follow-up method
used on the most significant outliers of the
Hough Einstein@Home all-sky search for continuous gravitational waves~\cite{Aasi:2012fw}.
This follow-up method is based on the two-stage approach introduced in
\cite{ShaltevPrix2013}, consisting of a semicoherent refinement
followed by a fully coherent zoom.
We quantify the efficiency of the follow-up pipeline using simulated
signals in Gaussian noise.
This pipeline does not search beyond first-order frequency spindown,
and therefore we also evaluate its robustness against second-order
spindown.
We present the details of the Hough Einstein@Home
follow-up~\cite{Aasi:2012fw} on three hardware-injected signals and on
the 8 most significant \candidates of unknown origin.
\end{abstract}

\pacs{XXX}

\maketitle

\section{Introduction}
\label{sec:intro}

The search for unknown sources of continuous gravitational waves (CWs) is
computationally bound due to the enormous parameter space that needs to
 be covered \cite{Brady:1998nj}. Advanced
semicoherent search techniques, such as \cite{PhysRevD.72.042004,2009PhRvL.103r1102P}, are typically used to identify interesting regions
 of the parameter space, which then require fully coherent follow-up studies in order
to confirm or discard potential CW candidates.  The
parameter space associated with these candidates is substantially  smaller
than the original search space. However, it is still large
enough to lead to a prohibitive computing cost, when data of order of months or
 years is analyzed fully coherently with a classical grid-based method \cite{Brady:1997ji}.
Therefore, an alternative follow-up method was developed, which
 combines the $\mF$-statistic \cite{Jaranowski:1998qm}\cite{cutler05:_gen_fstat} with
a Mesh Adaptive Direct Search (MADS) \cite{Audet04meshadaptive}
algorithm. This allows us to fully coherently examine long
data sets at a feasible computational cost \cite{ShaltevPrix2013}.

In the present work we describe how the  two-stage algorithm proposed in \cite{ShaltevPrix2013}
was adapted to follow up the most significant outliers in the Hough S5 Einstein@Home
search \cite{Aasi:2012fw}.
 We first validate the follow-up pipeline by performing Monte-Carlo
 studies. We inject and search for simulated CW signals added into
 simulated Gaussian noise data. Then we show how  the
search method was applied to  35 outliers identified in the
 Einstein@Home search, 27 of which
are associated with 3 simulated signals (hardware injections,
discussed in Sec.\ref{sec:hwi}).

The paper is organized as follows. In Sec.~\ref{sec:eah} we briefly recap
the  Einstein@Home all-sky search for periodic gravitational waves
in data from the fifth LIGO science run (S5). In Sec.~\ref{sec:pipe}
 we summarize the two-stage follow-up method and introduce
the search pipeline.
The efficiency of the follow-up algorithm is tested with
Monte-Carlo studies presented in Sec.~\ref{sec:mc}.
In Sec.~\ref{sec:hwi} we present the follow-up results for the 27
outliers associated with 3 hardware injections. In Sec.~\ref{sec:fup}
we show the results of the follow-up for the remaining CW outliers.
Section~\ref{sec:dis} presents a discussion of the results and
concluding remarks.

\subsubsection*{Notation and conventions}
When referring to a quantity $Q$ of the original Hough search, we
denote it as $Q_\HS$. A quantity measured after the pre-refinement
stage is denoted as $Q_\PR$, after the refinement as $Q_\R$, and as
$\co{Q}_\Zoom$  after the fully coherent zoom stage using all the
data (consistent with the notation of \cite{ShaltevPrix2013}).
We use an overbar ($\avg{Q}$) to denote an average over segments.

\section{The Hough S5 Einstein@Home all-sky search}
\label{sec:eah}

The Einstein@Home all-sky search \cite{Aasi:2012fw} uses the semicoherent Hough-transform method
 \cite{Krishnan:2004sv}, which consists of dividing the entire data span into $\Nseg$ shorter segments of duration $\Tseg$. In a first step, a coherent $\mF$-statistic search is performed
on a coarse grid for each of the data segments. Then the
Hough number-count statistic, defined in Eq.~\eqref{eq:7}, is computed on
 a finer grid, using the $\mF$-statistic values from the individual segments.

In this paper we focus on the S5R5 search of \cite{Aasi:2012fw}, which spans approximately $264$ days of data from
the Hanford (H1) and Livingston (L1) LIGO detectors. This dataset was
divided into $\Nseg=121$ segments of duration $\Tseg = 25$ hours each.
The parameter space covered by this search spans the entire sky, a frequency range
$f\in[50,1190]\ \Hz$, and a spindown range
$\dot{f}\in[-20,1.1]\times10^{-10}\ \Hz\ \seconds^{-1}$.

The phase evolution of the expected signal  at  the detector  can be written
 as \cite{Jaranowski:1998qm}
\begin{eqnarray}
\label{eq:1}
 \Phi(t)&\approx&\Phi_{0}
+ 2\pi\sum_{k=0}^{s}\frac{f^{(k)}(t_{0})(t-t_{0})^{k+1}}{(k+1)!}\\\nonumber
&+&2\pi\frac{\vec{r}(t)}{c}\vec{n}\sum_{k=0}^{s}\frac{f^{(k)}(t_{0})(t-t_{0})^{k}}{
k!}\ ,
\end{eqnarray}
where $\Phi_{0}$ is the initial phase,
 $f^{(k)}\equiv\frac{d^{k}f}{dt^{k}}$ represent the time
derivatives of the signal frequency $f$ at the solar system barycenter (SSB) at
reference time $t_{0}$, $s$ is the maximal considered spindown order, $c$ is the speed of
light, and $\vec{r}(t)$ is the vector pointing from the SSB to the
detector. The unit vector
$\vec{n}\equiv(\cos\alpha\cos\delta,\sin\alpha\cos\delta,\sin\delta)$
points from the SSB to the CW source, where $\alpha,\,\delta$ are the
standard equatorial coordinates referring to right-ascension and
declination, respectively.

 The $\mF$-statistic is one of the
 standard coherent techniques used to extract the CW signals from the
 noisy detector data. This statistic is the result of
 matched-filtering the data with a signal template
characterized by the phase-evolution parameters
 $\lambda \equiv\{\alpha,\delta,f,\dot{f}\}$.
The amplitude parameters, namely, the intrinsic amplitude
 $h_{0}$, the inclination angle $\iota$, the polarization angle $\psi$
and the initial phase  $\phi_{0}$ have been
analytically maximized over \cite{Jaranowski:1998qm}.
In a coherent
 grid-based
$\mF$-statistic search the number of templates increases with a high
power of the observation time \cite{prix06:_searc}.
 Hence these searches are not suitable for wide parameter-space all-sky surveys. However,
the  reduction of the coherent baseline in a semicoherent search
 \cite{Brady:1998nj,PhysRevD.72.042004} makes these techniques
computationally feasible in a distributed computing environment such as Einstein@Home, and (usually) more sensitive at fixed computing cost \cite{PrixShaltev2011}.

The template bank used to cover the parameter space is constructed
using the notion of mismatch \cite{bala96,owen96}. This is defined as
the fractional loss of squared signal-to-noise ratio (SNR) between a template $\lambda$ and the signal location $\lambda_\sig$.
We use the definition of SNR given in \cite{Jaranowski:1998qm}, and
denote it as $\snr$.

To quadratic order in parameter-space offsets
$\Delta\lambda^i\equiv\lambda^i - \lambda_\sig^i$, the mismatch can be approximated by
\begin{equation}
 \label{eq:3}
\mty\equiv g_{ij}(\lambda_\sig)\Delta\lambda^{i}\Delta\lambda^{j}\,,
\end{equation}
where $g_{ij}$ is a symmetric positive-definite matrix referred to as
the parameter-space metric. The indices $i,j$ label the
phase-evolution parameters, and we use summation convention over repeated indices.
This metric mismatch $\mty$ can be interpreted as a distance measure
in parameter space.

In the S5R5 analysis the templates at frequency $f$ were placed on a
coarse grid constructed using the following spacings
 \cite{Aasi:2012fw}:
\begin{equation}
\label{eq:4}
 d\theta_{\mF} = \frac{\sqrt{3}\,c}{v_{d}f\Tseg},\quad
df = \frac{\sqrt{12\,m}}{\pi\Tseg},\quad
d\dot{f} = \frac{\sqrt{3.3\,m}}{\Tseg^{2}}\,,
\end{equation}
where $d\theta_{\mF}$ is the angular resolution of the coarse sky
grid, $df,\,d\dot{f}$ are the frequency and spindown resolutions,
respectively; $m$ is the nominal single-dimension mismatch,
 taken equal to $0.3$ in \cite{Aasi:2012fw}, and
$v_{d}$ is the Earth's rotation speed at the equator.
Due to limitations of the Einstein@Home environment on the memory
footprint of the application, the spindown resolution was not
increased for the fine grid. Instead the $d\dot{f}$-resolution of
Eq.~\eqref{eq:4} was determined in a Monte-Carlo study so as to not significantly lose detection efficiency.

The resolution of the fine sky grid at frequency $f$ is given by \cite{Aasi:2012fw}
\begin{equation}
 \label{eq:5}
d\theta_{H} = \frac{c\,df}{\wp fv_{y}}\ ,
\end{equation}
where $\wp$ is the pixel factor and $v_{y}$ is the Earth's orbital velocity.  With
$\wp = 0.5$, $m = 0.3$ the sky refinement used in the S5R5 search yields
 $\Nsky^{\mathrm{ref}}=(d\theta_{\mF}/d\theta_{H})^{2}\approx8444$ \cite{Aasi:2012fw}.

Every parameter-space point of the search is assigned a significance,
or critical ratio $(\CR)$, value \cite{Aasi:2012fw}:
\begin{equation}
 \label{eq:6}
\CR = \frac{n_\crit-\bar{n}_\crit}{\sigma}\ ,
\end{equation}
with
\begin{equation}
 \label{eq:7}
n_\crit=\sum_{\ell=1}^{\Nseg}w_{\ell} \,n_{\ell}
\end{equation}
the Hough number count, where $w_{\ell}$ is the weight for segment
$\ell$ at a frequency  $f$ and a sky position $(\alpha,\delta)$; $n_{\ell}=1$ if the
$\mF$-statistic crosses a certain threshold value (namely $2\mF>5.2$ in \cite{Aasi:2012fw}) otherwise
 $n_{\ell}=0$;
 $\bar{n}_\crit$ and  $\sigma$ are the expected value and the standard
 deviation of $n_\crit$ in Gaussian noise.
The candidates are ordered by their significance.

\section{Follow-up method}
\label{sec:pipe}

\subsection{The modified two-stage follow-up}

A slightly adapted version of the two-stage follow-up procedure
\cite{ShaltevPrix2013} was used in \cite{Aasi:2012fw} and is presented
here.
As mentioned in Sec.~\ref{sec:eah}, the original Hough search did not
use refinement in $\dot{f}$ and this led to a reduction in localization accuracy.
To recover from this, we perform a \emph{pre-refinement} by re-running
the original Hough search with a finer grid around the outlier being followed up.
Namely, we increase the resolution of the $\dot{f}$-grid by a factor
$\Nseg=121$, and the sky-resolution by doubling the pixel factor $\wp$
in Eq.~\eqref{eq:5}.
The usefulness of this pre-refinement is illustrated in Fig.~\ref{fig:mismatch}.

The loudest parameter-space point after pre-refinement provides the starting point
for the subsequent MADS-based follow-up method described in \cite{ShaltevPrix2013}:
Namely, we first employ the semicoherent
$\mF$-statistic $\avg{2\mF}$, defined as
\begin{equation}
  \label{eq:13}
  \avg{2\mF}(\lambda) \equiv \frac{1}{\Nseg}\sum_{\ell=1}^{\Nseg} 2\mF_\ell(\lambda)\,,
\end{equation}
where $2\mF_\ell(\lambda)$ is the coherent $\mF$-statistic computed on
segment $\ell$ at the parameter-space point $\lambda$.
This is computed on the original Hough segments to further
improve the localization of the maximum-likelihood parameter-space
point, using the gridless MADS search method described in more detail
in \cite{ShaltevPrix2013}.
This stage is called \emph{refinement}, with detection statistic
$\avg{2\mF}_\R$ for the loudest resulting candidate.

Next we apply the so-called $\mF$-statistic consistency veto
of \cite{Aasi:2012fw,Keitel:2013wga}, namely
\begin{equation}
  \label{eq:2}
  \text{veto if }
  \avg{2\mF}_\R < \max\{\avg{2\mF}^{\HAN}_\R,\,\avg{2\mF}^{\LIV}_\R \}\,,
\end{equation}
where $\avg{2\mF}_\R^{\HAN,\LIV}$ denote the corresponding semicoherent $\mF$-statistic
values from the individual detectors H1 and L1, respectively.

Then, in the so-called \emph{zoom} stage, we compute the fully-coherent $\co{2\mF}_\Zoom$
statistic using all the data. From this we determine whether
the resulting candidate is consistent with the signal model or with
Gaussian noise.

\subsection{Classification of zoom outcomes}
\label{sec:class-zoom-outc}

We distinguish three possible outcomes of the zoom stage:
\begin{itemize}

 \item \textit{Consistency with Gaussian noise ($\CGN$)} - the fully coherent $\co{2\mF}_\Zoom$
value does not exceed a prescribed threshold, i.e.,
\begin{equation}
\label{eq:8}
 \co{2\mF}_\Zoom<\CGNth\,,
\end{equation}
where $\CGNth$ is chosen to correspond to some (small)
false-alarm probability $\pfA$ in Gaussian noise. The single
trial false-alarm probability for a given $\CGNth$ threshold
is $\pfA^{1} = (1+\mF)e^{-\mF}$ ( see, e.g.,  \cite{ShaltevPrix2013} for details).
For example, we find that a threshold of $\CGNth=90$
 corresponds to a false-alarm probability $\sim\mO(10^{-18})$ for a
 single template.  Assuming $\Ntemp$ independent templates and
 $\pfA\ll1$, the false-alarm is $\pfA\approx\Ntemp\pfA^{1}$.

\item \textit{Non-Gaussian origin ($\NGO$)} - the candidate is loud enough to be
inconsistent with Gaussian noise at the chosen $\pfA$, i.e.,
\begin{equation}
 \label{eq:9}
\co{2\mF}_\Zoom\ge\CGNth\ .
\end{equation}

\item \emph{Signal recovery} ($\SMC$) - defined as a \emph{subclass} of $\NGO$,
namely a signal is considered recovered if for the final zoomed candidate
the $\co{2\mF}_\Zoom$ value exceeds the Gaussian-noise threshold $\CGNth$
\emph{and} falls into a predicted signal interval:
\begin{equation}
 \label{eq:10}
  \coFth < \co{2\mF}_\Zoom < \coFmax\ ,
\end{equation}
where $\coFth \equiv \max\{ \CGNth, \, \co{2\mF}_o -
n_{u}\,\sigma_o\}$, and
$\coFmax\equiv \co{2\mF}_o + n_{u}\,\sigma_o$,
with expectation
\begin{equation}
\label{eq:11}
\co{2\mF}_o \approx 4 + \Nseg\,\left(\avg{2\mF}_\R - 4\right)\,,
\end{equation}
and variance
\begin{equation}
\label{eq:12}
\sigma^2_o \approx 2\left( 4+2\Nseg\,\left(\avg{2\mF}_\R - 4\right)\right)\,.
\end{equation}
The number $n_{u}$ determines the probability that a true signal
candidate would fall into this interval.
For example, $n_{u}=6$  corresponds roughly to a confidence
of$~\sim99.6\%$ (provided $\CGN$ and $\SMC$ are disjoint).

\end{itemize}

\subsection{Choice of MADS parameters}
\label{sec:mesh-adaptive-direct}

In both stages
the parameter space is explored on a dynamically created mesh by
using a MADS-based algorithm \cite{Audet04meshadaptive}. MADS itself
 is a general purpose
algorithm for derivative-free optimization, which is typically applied to
computationally expensive problems with unknown derivatives.
The input to the MADS-based algorithm is a starting point
$\lambda_\cand$, a search bounding box $\Delta\lambda_\R$ around
$\lambda_\cand$ and a set of MADS parameters, which govern the choice
of evaluation points, namely
$\{d\lambda,\mub,\minmce,\maxmce,\mre,\maxbbeval\}$, where
$d\lambda$ is the initial step, $\mub$ is the mesh update basis, $\minmce$
and $\maxmce$ are the mesh-coarsening exponents, $\mre$ denotes
the mesh-refining exponent and $\maxbbeval$ is the maximum number of
templates to search over; for details we refer the reader to Sec. IIIE
in \cite{ShaltevPrix2013}.
The algorithm parameters for the MADS-based refinement and zoom
stage are summarized in Table \ref{tab:mc-parameters}. These parameters have been found to yield good results in Monte-Carlo studies.
\begin{table}[htbp]
 \centering
\begin{tabular}[c]{c|c|c|c|c|c}
stage & $\mre$ & $\minmce$ & $\maxmce$ & $\mub$ & $\maxbbeval$\\\hline
R & -1 & 1 & 20 &  2 & 20000 \\\hline
Z & -1 & 1 & 50 &  1.2 & 20000
\end{tabular}
\caption{Follow-up algorithm parameters for the refinement and zoom stage.}
\label{tab:mc-parameters}
\end{table}

\subsection{Follow-up parameter-space regions}
\label{sec:follow-up-parameter}

We stress that the bounding box $\Delta\lambda_\R$ used for the
refinement differs with respect to what is described in \cite{ShaltevPrix2013}.
There the refinement is restricted to the semicoherent metric
ellipsoid centered on a candidate.
Here, instead, the refinement stage is performed on a box that was
empirically determined to be large enough to contain the true signal
location with very high confidence:
\begin{equation}
 \label{eq:15}
  \begin{split}
    \Delta \alpha = 0.4\ \rad\,,&\quad\Delta \delta = 0.4\ \rad\\
    \Delta f = 1\times10^{-4}\ \Hz\,,&\quad\Delta\dot{f} = 1\times10^{-9}\ \Hz/\seconds\ .
  \end{split}
\end{equation}
Given that this follow-up was not computationally limited, we did not
attempt to find the smallest possible refinement region.

The zoom search is constrained by a Fisher ellipse scaled to
24 standard deviations, as described in \cite{ShaltevPrix2013}.
This large number was chosen empirically by increasing it until the
pipeline performance did not further improve.

The minimal spindown order required is related to parameter-space
thickness measured in terms of the extent of the metric ellipse along
that direction \cite{Brady:1998nj,PhysRevD.72.042004,PrixShaltev2011}.
As a rule of thumb, the maximal spindown order required in a search
increases with the time spanned by the data.
In the Hough Einstein@Home all-sky search \cite{Aasi:2012fw}, the
follow-up procedure did not include second-order spindown.
In Secs.~\ref{sec:mc} we show the performance of
the follow-up pipeline on signals with zero second-order spindown,
while in Sec.~\ref{sec:mcspd} we study the robustness of this method
in the case of maximal second-order spindown (as considered in \cite{Aasi:2012fw}).

\section{Monte-Carlo studies}
\label{sec:mc}

\begin{figure*}[htbp]
\centering
\subfloat[ ]{\includegraphics[width=\columnwidth]{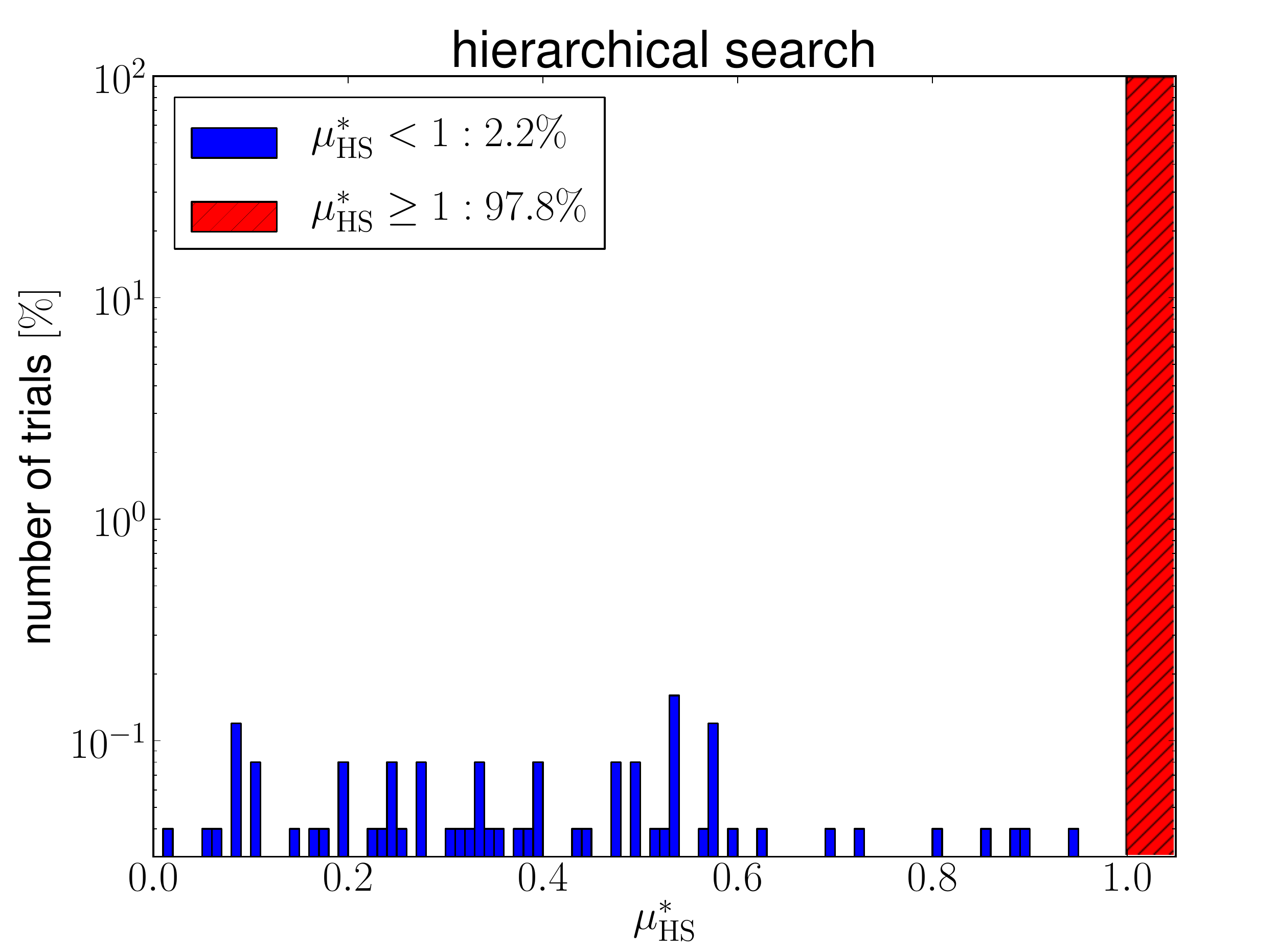}\label{fig:mis_a}}
\quad
 \subfloat[ ]
{\includegraphics[width=\columnwidth]{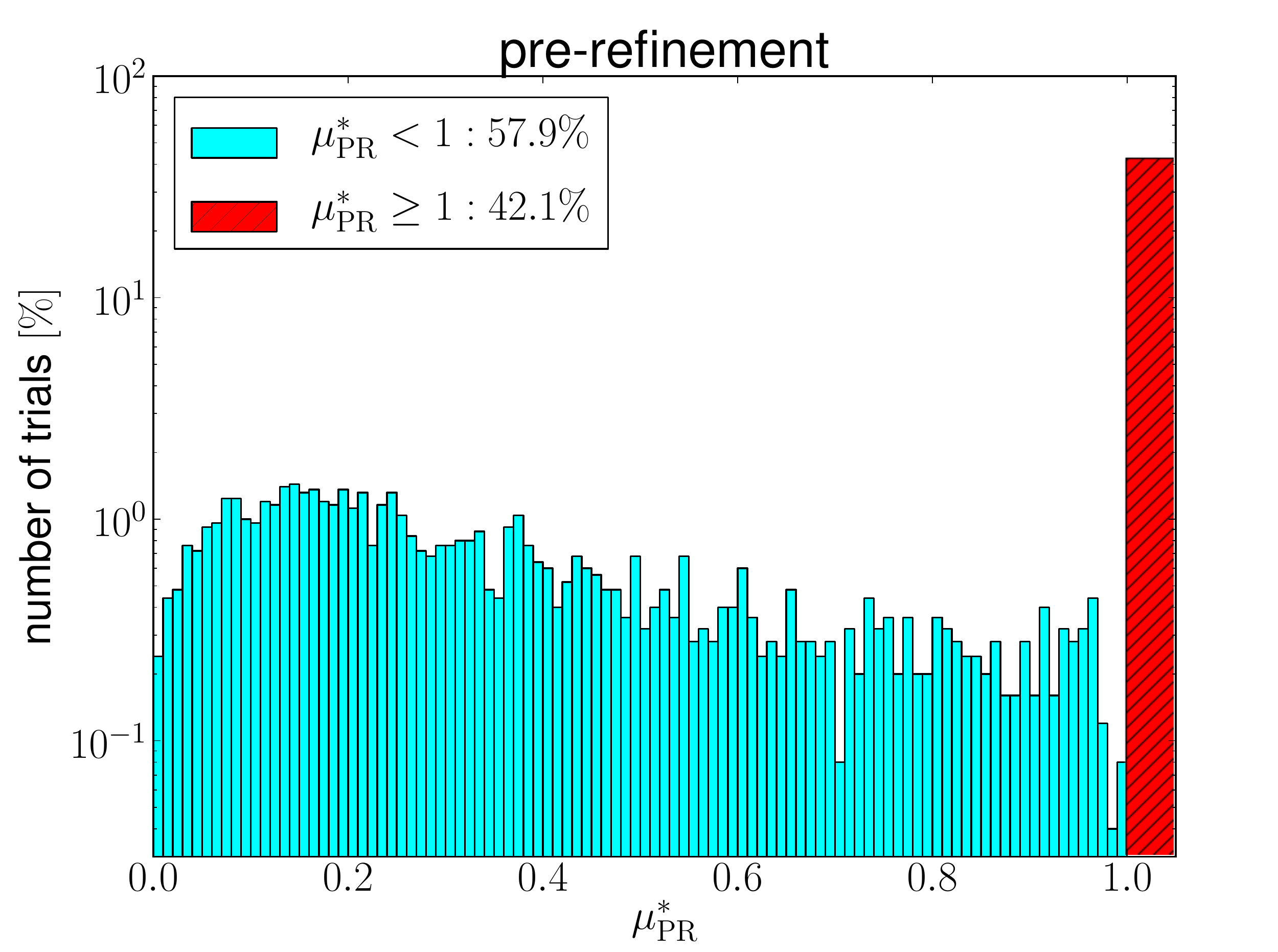}\label{fig:mis_b}}
\quad
 \subfloat[ ]
{\includegraphics[width=\columnwidth]{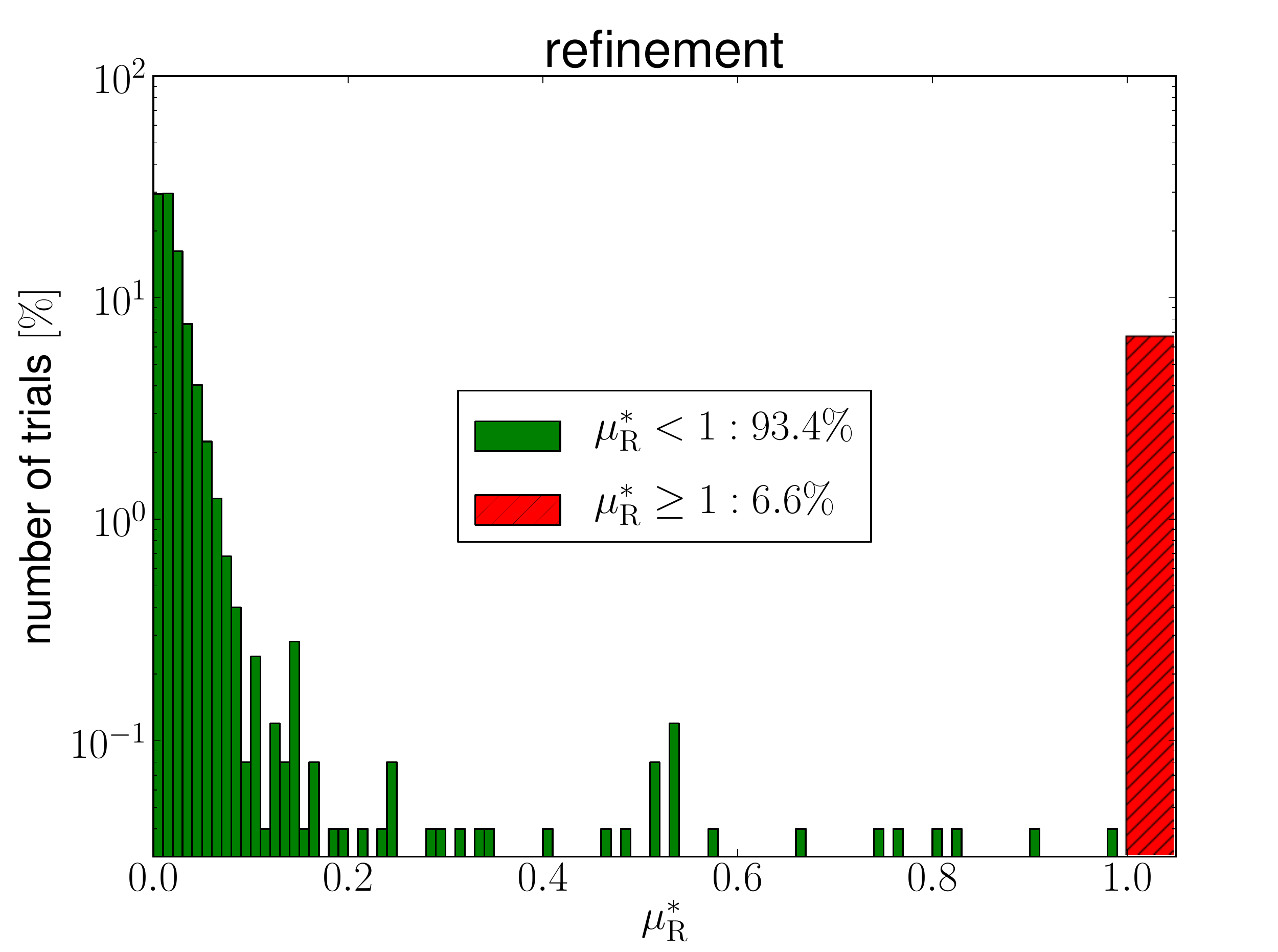}\label{fig:mis_c}}
\caption[Monte-Carlo study of the efficency of the follow-up pipeline.]
{Semicoherent metric mismatch  of the subset of 2500 injections with
$\avg{\snrsq}\in{[5,6]}$ at different stages of the Monte-Carlo.
The panel (a) shows the metric mismatch  distribution after the original
hierarchical search.  The panel (b) shows the metric mismatch
 distribution after the pre-refinement stage. The panel (c) shows the
 metric mismatch distribution after the refinement stage. The
hatched bar in each panel shows the percentage of trials
with $\mu^{*}\ge1$.}
\label{fig:mismatch}
\end{figure*}

\subsection{Setup}
\label{sec:setup}

We test the proposed follow-up pipeline in an end-to-end Monte-Carlo study using
the LALSuite \cite{LALSuite:Misc} software package. In particular we use the following
 LALApps applications: \texttt{Makefakedata\_v4} to generate Gaussian
noise and inject CW signals; \texttt{FstatMetric\_v2} to compute the
 fully coherent or semicoherent  metric;
\texttt{HierarchicalSearch} for the semicoherent Hough-transform
 search; \texttt{FStatSCNomad} for the semicoherent $\mF$-statistic optimization with
MADS, and \texttt{FStatFCNomad} for the fully coherent $\mF$-statistic MADS optimization,
where for the MADS algorithm we use the reference implementation NOMAD \cite{LeDigabel2011A909}.

\begin{figure}[htbp]
 \includegraphics[width=\columnwidth]{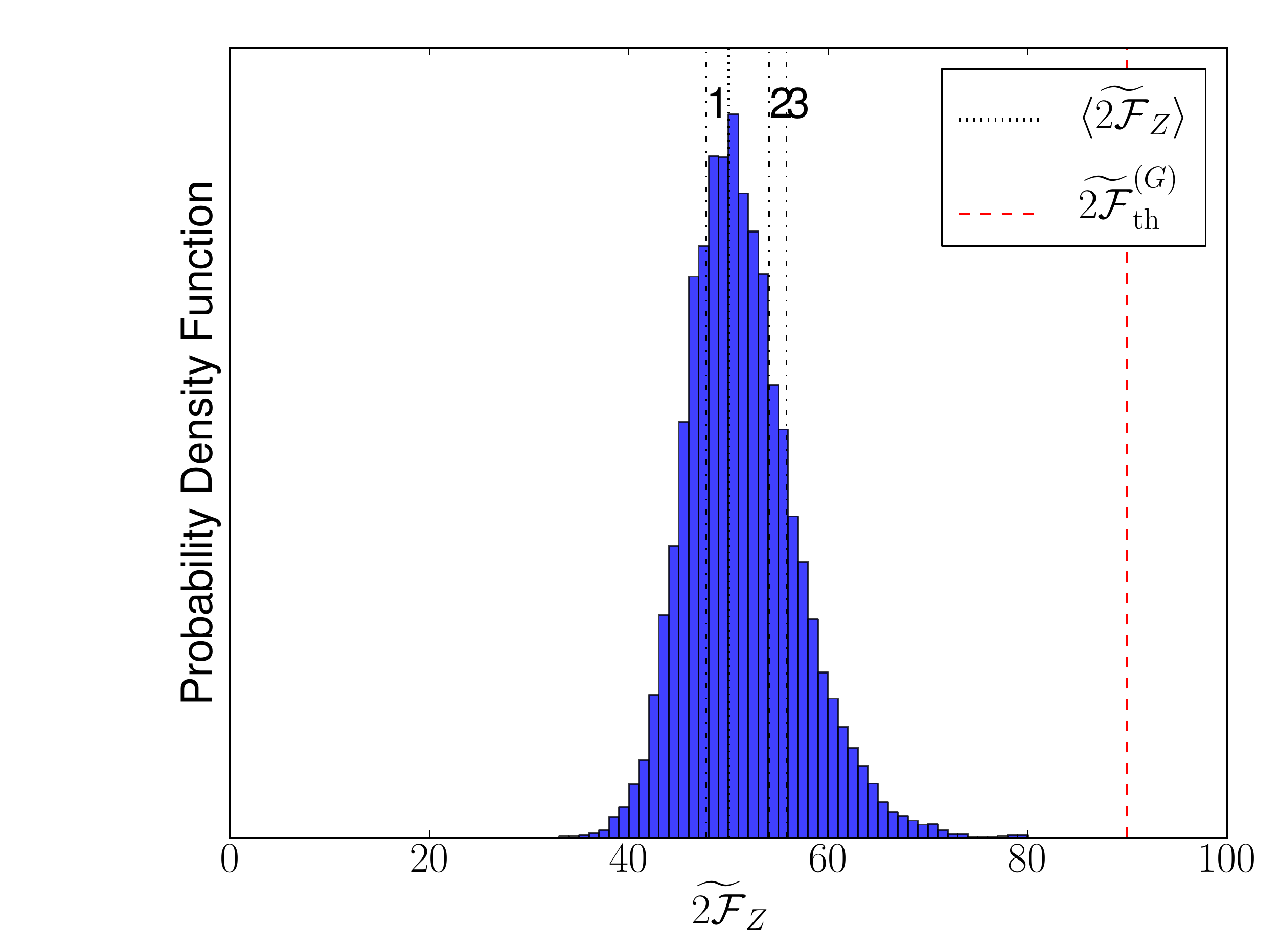}
\caption{The figure
shows the $\co{2\mF}_\Zoom$ distribution after the fully coherent 4-D $\{\alpha,\delta,f,\dot{f}\}$ zoom stage of
 15000 searches in pure Gaussian noise data without injected signals.
The maximum value is $\co{2\mF}_\Zoom^{\max}=$
 \protect\input{S5R5_noise_dist_max.tex}, and
 the mean value is $\tavg{\co{2\mF}_\Zoom}=$
 \protect\input{S5R5_noise_dist_avg.tex} (dotted line).
 The labels 1,2,3 refer to the \candidates at roughly $434$, $677$ and
 $984$ Hz, respectively (see Tables \ref{tab:discarded}).
 The vertical red line marks the noise threshold.}
\label{fig:detections_a}
\end{figure}
We apply the follow-up chain to 15000 different noise realizations with and
 without injected signals. The Gaussian noise realizations  are generated with
the \texttt{MakeFakedata\_v4} application using the same timestamps of the SFTs
\footnote{SFT is the acronym used for Short time baseline Fourier Transform of
 the calibrated detector strain data. The duration of the SFTs is typically
1800 seconds. SFTs are used as input to many CW searches such as the
semicoherent Hough-transform search, as well as the fully coherent follow-up.}
 used in the original Einstein@Home search with detector noise level
of $\sim2\times10^{-23}\,\Hz^{-1/2}$ per detector.
The signal parameters are uniformly drawn in the ranges $\cos\iota\in[-1,1]$,
$\psi\in[-\pi/4,\pi/4]$, $\phi_{0}\in[0,2\pi]$ and $f\in{[185,186]}\ \Hz$,
and the sky position is drawn isotropically on the sky. The frequency
range has been  chosen in the most sensitive region of the
LIGO detectors. The spindown value is randomly chosen in the range
 $\dot{f}\in(-\frac{f_{0}}{\tau_{0}},0.1\frac{f_{0}}{\tau_{0}})$
with minimal spindown age $\tau_{0}=800\ \years$ at $f_{0}=50\ \Hz$.
The signal amplitude is high enough such that the $\SNRSQ$ in the
point of injection is uniformly distributed in the range
 $\avg{\snrsq}\in[0,6]$.

We begin the end-to-end validation with a simulation stage of the original
S5R5 Einstein@Home search by using the original search setup, i.e., the
same frequency and spindown grid spacings given by Eq. \eqref{eq:4}.
The S5R5 search has been partitioned in independent computing
tasks, referred to as workunits (WUs). For a detailed discussion of the
WU see Sec. III C in \cite{Aasi:2012fw}. To save computing power,
we do not rerun an entire  WU in this simulation stage, but we center a search
grid around a  random point in the vicinity of the injected signal, searching over 10
frequency bins in total.  The  sky grid is constructed by extracting
16  points around the candidate from the original sky-grid file.
However, this reduced
parameter-space size is still sufficiently large to make possible the
selection of candidates due to the noise, if the signal is  weak as
might happen in  a real search, and not artificially  select a point
close to the true signal location.

In Fig. \ref{fig:mismatch} we show the semicoherent metric mismatch
distribution, computed with Eq. \eqref{eq:3}, after the Hough search,
the pre-refinement Hough search and after the refinement stage.
The loudest point selected from the refinement stage is
used as a starting point for the fully coherent $\mF$-statistic zoom search.

\subsection{Efficiency of the follow-up pipeline}
\begin{figure*}[htbp]
\centering
\subfloat[ ]{\includegraphics[width=\columnwidth]{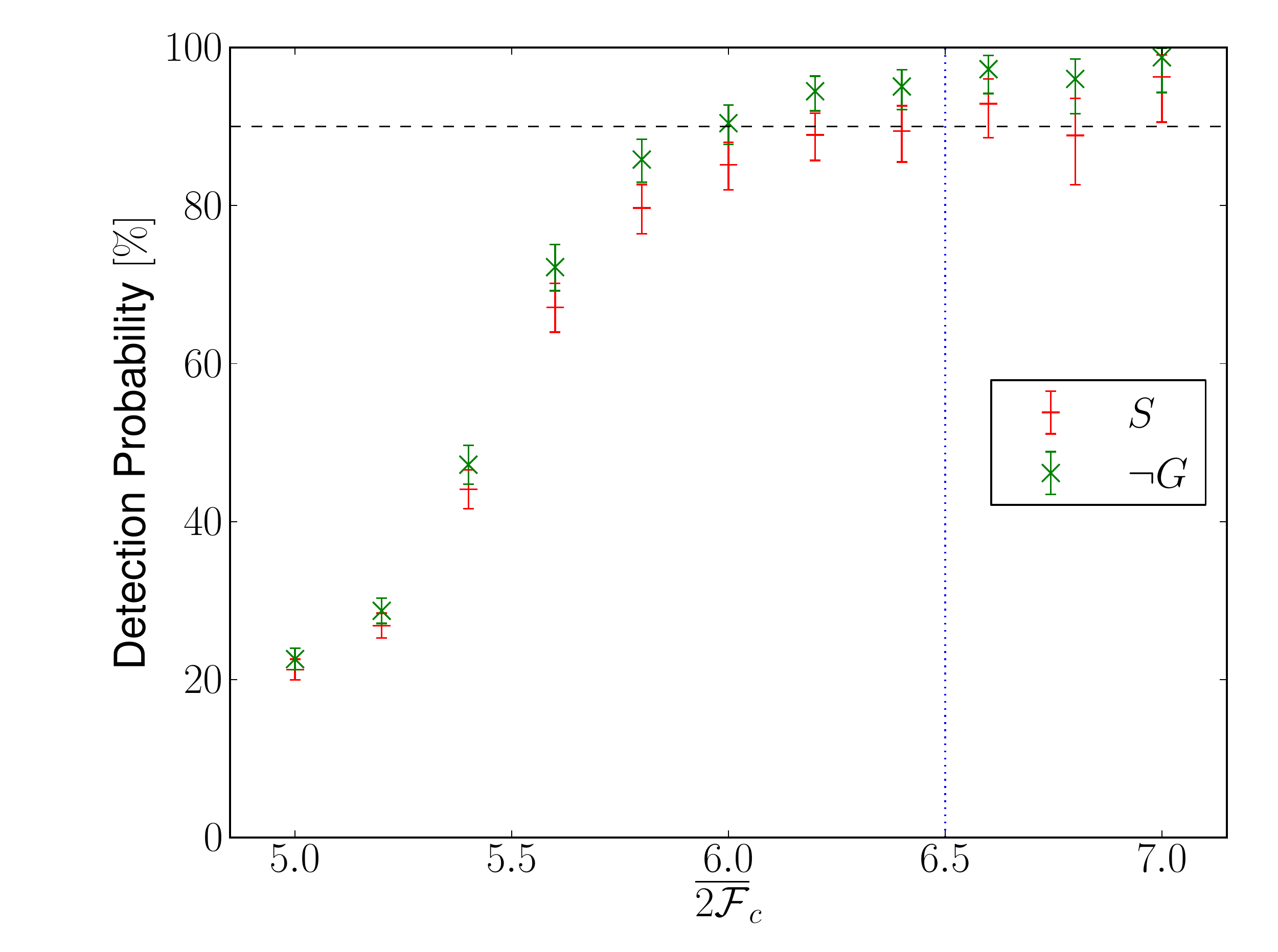}\label{fig:detections_b0}}
\quad
 \subfloat[ ]
{\includegraphics[width=\columnwidth]{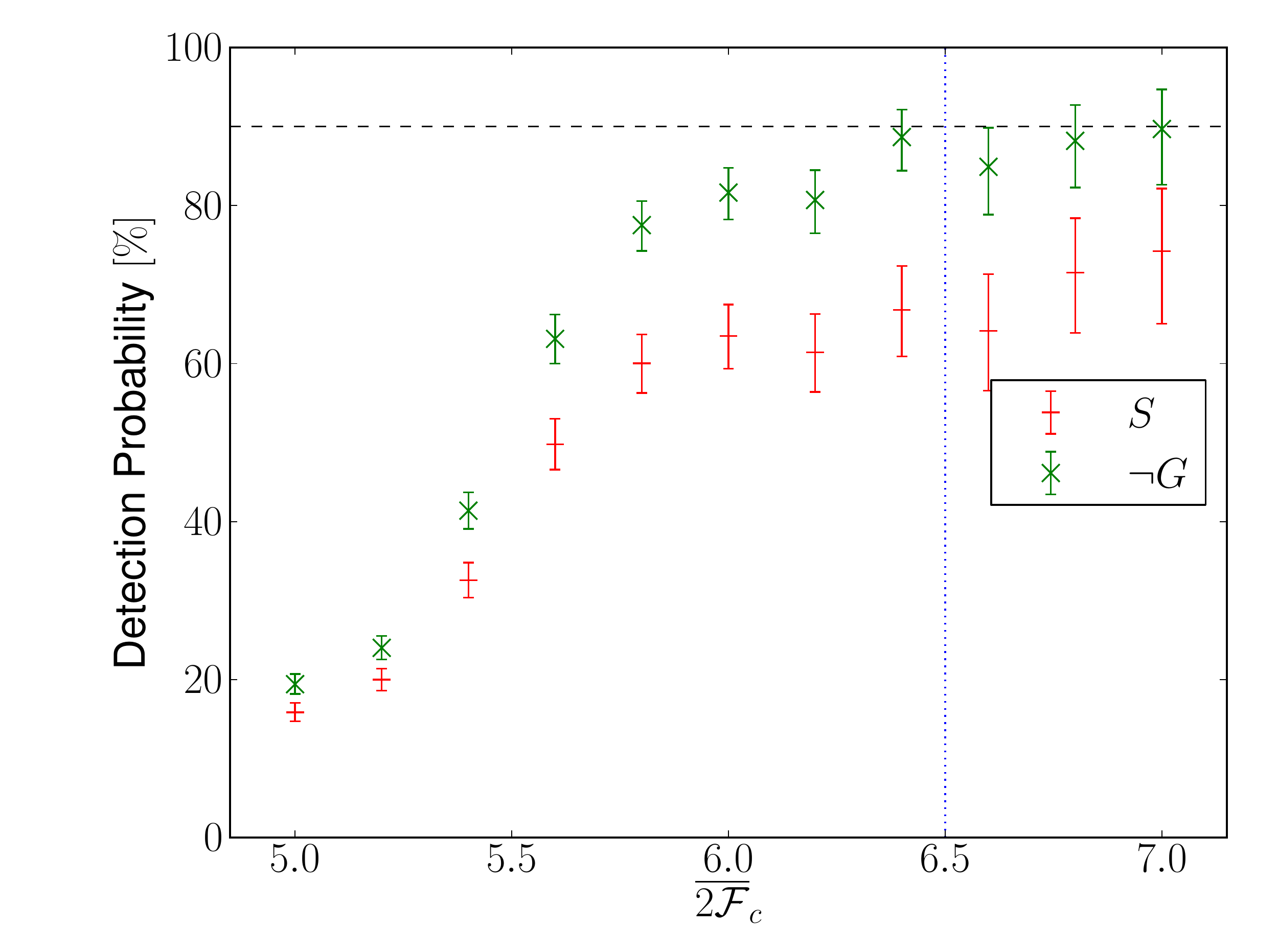}\label{fig:detections_b}}
\caption[Monte-Carlo study of the efficency of the follow-up pipeline.]
{ Monte-Carlo study of the efficiency of the follow-up pipeline. The panel (a) shows the percentage of the injected signals without second-order spindown classified as recovered (\textcolor{red}{$-$} $\ \SMC$),
and with a non-Gaussian origin ($\textcolor{green}{\times}\ \NGO$), as
function of the average $2\mF$ value of the candidate after the
original semicoherent Hough-transform search ($\avg{2\mF}_\cand$).
The panel (b) shows the percentage of the injected signals with second-order spindown classified as
 recovered (\textcolor{red}{$-$} $\ \SMC$),
and with a non-Gaussian origin ($\textcolor{green}{\times}\ \NGO$), as
function of $\avg{2\mF}_\cand$.
  The error bars account for $95\%$ confidence level. The $90\%$ detection probability is marked with the horizontal dashed line. The vertical dotted line denotes the $2\mF$
 threshold used to select candidates in \cite{Aasi:2012fw}.}
\label{fig:detections}
\end{figure*}

We first apply the follow-up pipeline to Gaussian noise data without any
injected signals. This is required to ensure
the applicability of the threshold $\CGNth = 90$  used
to consider a candidate as conform with the Gaussian noise hypothesis.
The distribution of the $\co{2\mF}_\Zoom$ values is plotted in
Fig.~\ref{fig:detections_a}.
The maximal value found is
$\co{2\mF}_\Zoom^{\max}=\input{S5R5_noise_dist_max.tex}$, which
is well below the $\NGO$ treshold of $\CGNth=90$.

In Fig. \ref{fig:detections_b0}
we plot the percentage of the injected signals classified as recovered ($\SMC$),
and as of non-Gaussian origin ($\NGO$), as a function of the average $2\mF$ value of the
candidate after the simulation stage. We are able to distinguish
$\ge 90\,\%$ of the candidates from Gaussian noise above
$\avg{2\mF}_\cand \gtrsim 6.0$, and we recover $\ge 90\,\%$
of the signals ($\SMC$)
for candidates with $\avg{2\mF}_\cand \gtrsim 6.2$.

\subsection{Robustness to second-order spindown signals}
\label{sec:mcspd}
The follow-up pipeline described in this work is limited to
first-order spindown in the signal model, which can lead to losses of
SNR over long observation times for signals with nonzero
second-order spindown. In order to test the robustness of the
follow-up method, we repeat the Monte-Carlo simulation on signals with
a fixed second-order spindown
value of $\ddot{f}=8\times10^{-20}\,\Hz/\seconds^{2}$, which corresponds to
the maximum considered in \cite{Aasi:2012fw}.
The result of this simulation is presented in Fig. \ref{fig:detections_b},
and shows that for candidates with $\avg{2\mF}_{\cand}\approx6.5$ the
follow-up pipeline is still able to distinguish close to $90\%$
of the candidates from Gaussian noise. Given that this was the
detection threshold used in the S5R5 search \cite{Aasi:2012fw}, we
conclude that the follow-up of the resulting candidates did not substantially
reduce the detection efficiency of the original search.

\section{Follow up of S5R5 \candidates}
\label{sec:follow-up-s5r5}

In this section we report details on the follow-up of the
S5R5 \candidates above the detection
threshold of $\avg{2\mF}\ge6.5$, as originally reported in \cite{Aasi:2012fw}.
For practical purposes these \candidates were divided into two classes,
depending on whether or not they are associated with hardware injections.

\subsection{\Candidates associated with hardware injections}
\label{sec:hwi}

\begin{table*}[htbp]
\centering
\begin{tabular}{|c||c||c||c|}\hline
Fake Pulsar & Pulsar 2 & Pulsar 3 & Pulsar 5 \\\hline\hline
$f_\sig\ [\Hz]$ & 575.16355763140 & 108.857159397497 & 52.8083243593 \\\hline
$\alpha_\sig\ [\rad]$ & 3.75692884 & 3.11318871 & 5.28183129 \\\hline
$\delta_\sig\ [\rad]$ & 0.06010895 & $-0.58357880$ & $-1.46326903$ \\\hline
$\dot{f}_\sig\ [\Hz/\seconds]$ & $-1.37\times10^{-13}$ & $-1.46\times10^{-17}$ &
$-4.03\times10^{-18}$\\\hline\hline
$\avg{2\mF}_{\HS}$ & 28 & 339 & 6.3 \\\hline\hline
$\avg{2\mF}_\R$ & 100 & 1137 & 12 \\\hline
$\avg{2\mF}^{\HAN}_\R$ & 51 & 641 & 8.2 \\\hline
$\avg{2\mF}^{\LIV}_\R$ & 54 & 510 & 8.0 \\\hline
$\mty_\R$ & $4.01\times10^{-4}$ & $5.18\times10^{-4}$ &
$4.88\times10^{-3}$ \\\hline\hline
$f_\Zoom\ [\Hz]$ & 575.16355763214 & 108.857159397523 & 52.8083243548\\\hline
$\alpha_\Zoom\ [\rad]$ & 3.75692887 & 3.11318900 & 5.28181148
\\\hline
$\delta_\Zoom\ [\rad]$ & 0.06010925 & $-0.58357884$ &
$-1.46326569$
\\\hline
$\dot{f}_\Zoom\ [\Hz/\seconds]$ & $-1.37\times10^{-13}$& $3.30\times10^{-16}$ &
$1.85\times10^{-15}$\\\hline
$\co{2\mF}_\Zoom$ & 7399 & 87097 & 678 \\\hline
$\co{2\mF}^{\HAN}_\Zoom$ & 3519 & 47572 & 350 \\\hline
$\co{2\mF}^{\LIV}_\Zoom$ & 3896 & 39557 & 332 \\\hline
$\co{2\mF}_\sig$ & 7377 & 86968 & 677\\\hline
$\mty_\Zoom$ & $2.4\times10^{-3}$ & $1.4\times10^{-3}$ &
$6.7\times10^{-3}$ \\\hline
$\Delta f_\Zoom\,[\Hz]$ & $7.44\times10^{-10}$ & $2.58\times10^{-11}$ &
$-4.55\times10^{-9}$\\\hline
$\Delta\dot{f}_\Zoom\,[\Hz/\seconds]$ & $-4.33\times10^{-16}$ & $3.44\times10^{-16}$ &
$1.85\times10^{-15}$ \\\hline
$\Delta\gamma_\Zoom[\rad]$ & $2.97\times10^{-7}$ & $2.48\times10^{-7}$ &
$3.96\times10^{-6}$ \\\hline
\end{tabular}
\caption{Most significant outlier after follow-up for each of the
  three fake pulsars.
  The injected signal parameters are $f_\sig, \alpha_\sig,
  \delta_\sig,\dot{f}_\sig$.
  The value of the $\mF$-statistic at the injection point is
  denoted as $\co{2\mF}_\sig$.
  The localization error of the final outlier in frequency and
  spindown is $\Delta f_\Zoom$ and $\Delta \dot{f}_\Zoom$,
  respectively, and $\Delta\gamma_\Zoom = 
\arccos(\vec{n}_\cand\vec{n}_\sig)$ denotes the angular separation.}
\label{tab:hwi}
\end{table*}
The CW hardware injections (referred to as ``fake pulsars'') are simulated signals, physically added into the
control system of the interferometer to produce a detector response similar
to what should be generated if a CW is present.
The aim of such injections is to test and validate analysis codes and search pipelines.

The S5R5 Einstein@Home search \cite{Aasi:2012fw} identified three fake
pulsars, referred to as Pulsar 2, 3 and 5. In this section we detail
the follow-up of the \candidates associated with these hardware injections.
Each injection typically produced many significant outliers.
We apply a simple clustering algorithm in order to follow up only the
most interesting ones.
Namely, for each hardware injection, we identify the loudest outlier
and remove all neighboring \candidates falling into the refinement box
given in Eq. \eqref{eq:15}.
We repeat this procedure until there are no more \candidates left.
A similar clustering algorithm was developed for the galactic-center
search \cite{Aasi:2013jya_BB,BBThesis}.

There are, for instance, $88$ parameter-space points associated with Pulsar 2
injected at $\sim575\,\Hz$. After the clustering procedure, the number of
\candidates to follow up  is reduced to 16. For Pulsar 3, injected at $\sim108\,\Hz$,
the number of parameter-space points to follow up shrinks from 80 to 9.
For Pulsar 5, injected at $\sim52\,\Hz$, there are only 2 \candidates,
which fall into different search boxes and are therefore unaffected by
the clustering.

In Table \ref{tab:hwi} we summarize, for each fake pulsar, the
recovered parameters of the loudest outlier resulting from the
follow-up.
All the injections were recovered at parameter-space points very close
to the injected signal parameters, as quantified by the values of the
metric mismatch $\mty_\Zoom$.
We note that the recovered detection statistic $\co{2\mF}_\Zoom$ is
slightly above the value at the injection point $\co{2\mF}_\sig$,
which is generally expected to be true for the maximum, due to noise
fluctuations.

\subsection{\Candidates of unknown origin}
\label{sec:fup}

The S5R5 search additionally yielded 8 \candidates of
unknown origin above $\avg{2\mF}\ge6.5$.
The results of the follow-up are summarized in Table~\ref{tab:discarded}.
None of these \candidates were found to be consistent with the signal
hypothesis in the sense of Eq.~\eqref{eq:10}: either they failed the
$\mF$-statistic consistency veto of Eq.~\eqref{eq:2} after refinement,
or they were found to be consistent with Gaussian noise (in the sense
of Eq.~\eqref{eq:8}) after the zoom stage.

These \candidates, with frequencies at approximately 434, 677, and 984
$\Hz$, are shown in Fig.~\ref{fig:detections_a} against the
distribution of $\co{2\mF}_\Zoom$ values obtained in Gaussian noise.

\begin{table*}[htbp]
\centering
\begin{tabular}{|c||c|c|c||c||c|c|c||c|}\hline
 $f\ [\Hz]$& $\avg{2\mF}_\R$ & $\avg{2\mF}_\R^{\HAN}$ & $\avg{2\mF}_\R^{\LIV}$ & $\mF$-veto & $\co{2\mF}_\Zoom$ & $\co{2\mF}_\Zoom^{\HAN}$ & $\co{2\mF}_\Zoom^{\LIV}$&outcome\\\hline\hline
52 & 12 & 8.2 & 8.0 & pass & 678 & 350 & 332 & $S$ \\\hline
96 & 9.1 & 4.4 & 13 & fail & - & - & - & - \\\hline
108 & 1137 & 641 & 510 & pass & 87097 & 47572 & 39557 & $S$\\\hline
144 & 11 & 4.5 & 14& fail & - & - & - & - \\\hline
434 &  5.5 & 5.4 & 4.5 & pass & 47 & 30 & 22 & $G$\\\hline
575 & 100 & 51 & 54 & pass & 7399 & 3519 & 3896 & $S$\\\hline
677 & 6.4 & 5.4 & 5.2 & pass & 54 & 44& 14 & $G$\\\hline
932 & 7.6 & 8.0 & 4.2 & fail & - & - & - & - \\\hline
984 & 6.5 & 4.8 & 5.5 & pass & 55 & 36 & 20 & $G$\\\hline
1030 & 7.4 & 8.3 & 4.5 & fail & - & - & - & -  \\\hline
1142 & 8.5 & 10 & 4.2 & fail & - & - & - & - \\\hline
\end{tabular}
\caption[Follow-up results for the nine \candidates from the
S5R5 search.]{Summary of the follow-up results for the 3 loudest
  \candidates associated with hardware-injections and the 8 most
  significant remaining outliers from the S5R5 search.
  The last column gives the classification of the final outlier
  after zoom, provided it passed the $\mF$-statistic consistency veto.}
\label{tab:discarded}
\end{table*}

\section{Discussion}
\label{sec:dis}

In this paper we describe the extension of the two-stage follow-up
method of \cite{ShaltevPrix2013} that was developed in order to follow
up \candidates from the Hough S5 Einstein@Home all-sky search~\cite{Aasi:2012fw}.
The extension consists of an additional Hough search as a
pre-refinement step, and an $\mF$-statistic consistency veto after
refinement to reduce the false-alarm rate on real detector data.
Pre-refinement was found to be necessary to improve the localization
accuracy of the original search outliers.

With a Monte-Carlo study we quantify the detection probability as a
function of the initial candidate strength, as shown in
Fig.~\ref{fig:detections}.
In particular, we find that the pipeline achieves a detection
probability of $90\,\%$ for candidates without second-order
spindown at a strength of $\avg{2\mF}_\cand \gtrsim 6$.
On the other hand, for signals with maximal second-order spindown
(as considered by the original Hough Einstein@Home
search~\cite{Aasi:2012fw}), the detection efficiency is reduced: for
example, at $\avg{2\mF}_\cand \approx 6.5$ the probability of signal
recovery drops to $\approx 60\,\%$, while the pipeline is still able
to separate $\approx 90\,\%$ of injected signals from Gaussian noise.

We illustrate the performance of this pipeline on real data by
detailing the follow-up of Hough Einstein@Home \candidates, which was
first presented in \cite{Aasi:2012fw}.
The pipeline successfully detects the three hardware injections
present in the \candidates set and recovers their parameters with
high accuracy, see Table~\ref{tab:hwi}.
The follow-up of the 8 most significant \candidates of unknown origin
finds them to be consistent with either Gaussian noise or with line
disturbances in the data.

\section{Acknowledgments}
We are thankful for numerous discussions and comments from colleagues,
in particular Badri Krishnan, Alicia Sintes, Bruce Allen, David Keitel, Karl Wette and Stephen Fairhurst.  We are grateful
to Peter Shawhan, Teviet Creighton and Andrzej Krolak for comments on this work in the process of review
of \cite{Aasi:2012fw}.

MS gratefully acknowledges  the support of Bruce Allen and the IMPRS on
Gravitational Wave Astronomy of the Max-Planck-Society.
PL and MAP acknowledge support by the ``Sonderforschungsbereich''
Collaborative Research Centre (SFB/TR7).
This paper has been assigned AEI preprint number AEI-2014-009 and LIGO document number \dcc.

\bibliography{ppfupeah}
\end{document}

%% file: tag.tex
\def\commitID{commitID: 9234ee1101fc55c0aef83eb0518deb77315bfb6d}
\def\commitDATE{ Sun May 4 10:03:01 2014 +0200}
\def\commitIDshort{commitID: 9234ee1}
\def\commitSTATUS{CLEAN}

%% file: ppfupeahdefs.tex
\newcommand{\sig}{{\mathrm{s}}}

\newcommand{\mF}{\mathcal{F}}

\newcommand{\mO}{\mathcal{O}}

\newcommand{\Ntemp}{\mathcal{N}}
\newcommand{\Nsky}{\Ntemp_{\mathrm{sky}}}

\newcommand{\Tseg}{\Delta T}

\newcommand{\Nseg}{N}

\renewcommand{\vec}[1]{\mathbf{#1}}

\newcommand{\snr}{\rho}
\newcommand{\snrsq}{\snr^{2}}
\newcommand{\SNRSQ}{\mathrm{SNR}^{2}}

\newcommand{\CR}{\mathrm{CR}}
\newcommand{\crit}{{\mathrm{c}}}

\newcommand{\HS}{{\mathrm{HS}}}
\newcommand{\PR}{{\mathrm{PR}}}
\newcommand{\R}{{\mathrm{R}}}
\newcommand{\Zoom}{{\mathrm{Z}}}

\newcommand{\cand}{{\mathrm{c}}}

\renewcommand{\th}{\mathrm{th}}

\newcommand{\maxbbeval}{p}

\newcommand{\Hz}{\mathrm{Hz}}

\newcommand{\HAN}{\mathrm{H1}}
\newcommand{\LIV}{\mathrm{L1}}

\newcommand{\mub}{u_{b}}
\newcommand{\mce}{w^{+}}
\newcommand{\mre}{w^{-}}
\newcommand{\minmce}{\mce_{\mathrm{min}}}
\newcommand{\maxmce}{\mce_{\mathrm{max}}}

\newcommand{\coFth}{\co{2\mF}_{\mathrm{\th}}^{(\SMC)}}
\newcommand{\coFmax}{\co{2\mF}_{\mathrm{\max}}^{(\SMC)}}
\newcommand{\CGNth}{\co{2\mF}_{\mathrm{\th}}^{(\CGN)}}
\newcommand{\pfA}{p_{\mathrm{fA}}}

\newcommand{\seconds}{\mathrm{s}}

\newcommand{\years}{\mathrm{yr}}

\newcommand{\avg}[1]{\overline{#1}}
\newcommand{\tavg}[1]{\langle#1\rangle}

\newcommand{\rad}{\mathrm{rad}}

\newcommand{\mty}{\mu^{*}}

\newcommand{\co}[1]{\widetilde{#1}}

\newcommand{\SMC}{S}
\newcommand{\CGN}{G}
\newcommand{\NGO}{\neg\CGN}

\newcommand{\candidates}{search outliers\ }
\newcommand{\Candidates}{Search outliers\ }

%% file: S5R5_noise_dist_avg.tex
$51.40$

%% file: S5R5_noise_dist_max.tex
79.36